\documentclass[aps,pra,reprint,showpacs,notitlepage,superscriptaddress,twocolumn]{revtex4-2}
\usepackage{amssymb}
\usepackage{mathrsfs}
\usepackage{amsfonts}
\usepackage{graphicx}
\usepackage{amsmath}
\usepackage{dcolumn}
\usepackage{color}
\usepackage{subfigure}
\usepackage{epsfig}
\usepackage{soul}
\usepackage{upgreek}
\usepackage{graphicx} 
\usepackage[colorlinks,linkcolor=blue,citecolor=blue,hyperindex,bookmarks=false,pdfstartview=FitH]{hyperref}

\providecommand{\U}[1]{\protect\rule{.1in}{.1in}}

\newcommand{\figpanel}[2]{\hyperref[#1]{\ref*{#1}(#2)}}

\begin{document}

\title{Nonreciprocal Entanglement by Dynamically Encircling a Nexus}

\author{Lei Huang}
\affiliation{College of Physics and Electronic Engineering, Hainan Normal University, Haikou 571158, China}
\author{Peng-Fei Wang}
\affiliation{College of Physics and Electronic Engineering, Hainan Normal University, Haikou 571158, China}
\author{Jian-Qi Zhang}
\email{changjianqi@gmail.com}
\affiliation{Wuhan Institute of Physics and Mathematics, Innovation Academy of Precision Measurement Science and Technology, Chinese Academy of Sciences, Wuhan 430071, China}
\author{Xin Zhou}
\affiliation{College of Intelligence Science and Technology, National University of Defense Technology, Changsha 410073, China}
\author{Shuo Zhang}
\affiliation{Henan Key Laboratory of Quantum Information and Cryptography, Zhengzhou 450000, China}
\author{Han-Xiao Zhang}
\affiliation{College of Physics and Electronic Engineering, Hainan Normal University, Haikou 571158, China}
\author{Hong Yang}
\affiliation{College of Physics and Electronic Engineering, Hainan Normal University, Haikou 571158, China}
\author{Dong Yan}
\email{yand@hainnu.edu.cn}
\affiliation{College of Physics and Electronic Engineering, Hainan Normal University, Haikou 571158, China}

\date{\today}

\begin{abstract}
Nonreciprocal entanglement, characterized by inherently robust operation, is a cornerstone for quantum information processing and communications. However, it remains challenging to achieve nonreciprocal entanglement characterized by stability and robustness against environmental fluctuations. Here, we propose a universal nonlinear mechanism to engineer magnetic-free nonreciprocity in dissipative optomechanics by utilizing bistability, a phenomenon ubiquitous across nonlinear physical systems. By dynamically encircling the nexus of bistability, a cusp converged by the bistable surfaces, we obtain nonreciprocal displacement and then utilize it to achieve robust nonreciprocal entanglement. Owing to the unique landscape of bistability, our nonreciprocal displacement and entanglements exhibit stability and robustness through closed-loop operations. Our work presents a foundational framework for leveraging nonlinearity to achieve nonreciprocal quantum information processing. It paves new avenues for exploring nonreciprocal quantum information processing and designing backaction-immune quantum metrology with nonlinearity.
\end{abstract}

\maketitle

{\it Introduction:} Nonlinear dynamics, accurately describing the intricacies of real-world phenomena, is naturally present in various branches of science~\cite{may1976simple,bonan2008forests,boccaletti2006complex}. In biology, nonlinear dynamics govern complex behaviors such as chaotic neural networks~\cite{degn2013chaos}, stable ecosystems~\cite{may1977thresholds}, synchronization and rhythmic processes in physiology~\cite{glass2001synchronization}.
In chemistry, they influence oscillating reactions~\cite{petrov1997resonant} and pattern formations~\cite{vanag2000oscillatory,sakurai2002design}. 
In physics, nonlinear dynamics plays a crucial role in understanding and explaining complex behaviors and nonlinear phenomena in diverse physical systems, ranging from hydrodynamic systems~\cite{trefethen1993hydrodynamic,cross1993pattern,gavassino2024infinite,supekar2023learning}, to optical systems~\cite{boyd2008nonlinear,wang2023image,yildirim2024nonlinear,zhang2023graphene}, and then to quantum systems~\cite{peyronel2012quantum,chang2014quantum,roberts2020driven,yesharim2023direct,wang2023quantum,gao2023quantum}, each exhibiting nonlinear characteristics essential for the advancement of theoretical and experimental approaches in the discipline. One of the significant manifestations of nonlinear behaviors in physics is bistability~\cite{pisarchik2014control,bachtold2022mesoscopic}, in which a system can reside in either of two stable equilibrium states.

Bistability is not only a fundamental characteristic of classical nonlinear systems~\cite{shen2022mechanical} but also offers substantial potential for advancement in quantum information science~\cite{roberts2020driven}, owing to this unique nonlinear characteristic widely presented in various physical systems~\cite{zhou2023higher,pal2023programmable}. 
So far, bistability has been applied in various physical fields. In optics, the robustness of bistability ensures its applications in memory storage and signal processing~\cite{liu2010ultra,nozaki2012ultralow}, facilitating developments in communication technologies. In precision measurement, combined with additional noise, bistability can detect and amplify signals covered by environmental noises, resulting in stochastic resonance sensors~\cite{dykman1998can,gammaitoni1998stochastic,peters2021extremely,yuan2024stochastic} and generations of frequency combs~\cite{cingoz2012direct,gaeta2019photonic}. In topology, it predicts that the bistability of topological states would show their applications in topological quantum information processing~\cite{kartashov2017bistable}. In phase transition, bistability shows its power in switching states of matter~\cite{worrell2018bistable}. In optomechanics, it induces various physical phenomena~\cite{yang2023nonlinearity} including chaos~\cite{bakemeier2015route}, phonon laser~\cite{zhang2018phonon}, solitons~\cite{zhang2021optomechanical}, and entanglement~\cite{ghobadi2011quantum}. While bistability has broad applicability, its inherent link to symmetry breaking has yet to be exploited in the design of nonreciprocal quantum information processing. 
 
Nonreciprocity, characterized by the transmission of a signal depending on its propagation direction, is one of the main results of symmetry breaking~\cite{caloz2018electromagnetic}. 
It is an essential resource for exploring various applications~\cite{verhagen2017optomechanical}, including nonreciprocal sensing~\cite{lau2018fundamental}, isolators~\cite{ruesink2016nonreciprocity,peterson2017demonstration,miri2017optical,norte2016mechanical,bernier2017nonreciprocal}, circulators~\cite{ruesink2018optical,shen2018reconfigurable,barzanjeh2017mechanical}, unidirectional amplifiers~\cite{shen2018reconfigurable,bernier2017nonreciprocal,fang2017generalized}, and other vital components~\cite{shen2016experimental} for efficient information processing and communication networks.

As distinctive physical mechanisms can give rise to nonreciprocity in various physical systems, to date, we can categorize these physical mechanisms into the following main groups: (i) magnetic field-induced effects~\cite{haldane2008possible}, which contain magneto-optic effects~\cite{shayegan2022nonreciprocal} and topological materials~\cite{zhang2021superior}, (ii) dynamical modulations, such as time modulation~\cite{xu2016topological,doppler2016dynamically,abbasi2022topological,chen2022decoherence} and metamaterials~\cite{sounas2013giant,popa2014non,nassar2020nonreciprocity}, (iii) nonlinear processes~\cite{white2023integrated,wang2025self}, including scattering effects~\cite{dong2015brillouin} and asymmetric coupling~\cite{zhang2020breaking}. However, these physical mechanisms often face challenges in integration and require precise manipulations. Therefore, it is crucial to explore new physical mechanisms to overcome the constraints  and advance beyond current methods.

In this letter, we propose a new mechanism to generate magnetic-free nonreciprocity in a dissipatively coupled optomechanical system~\cite{xuereb2011dissipative,primo2023dissipative} by utilizing optomechanical bistability. We show that dynamically encircling a nexus of bistability gives rise to nonreciprocity. The critical principle is that closed trajectories encircling the nexus of bistability, involving transitions at boundaries of the bistable regime, result in direction-dependent outcomes when the start points are located in the bistable regime. It overcomes the loss of chirality that is associated with exceptional points~\cite{hassan2017dynamically,gao2025photonic,bu2024chiral,zhang2022topological} that arises from the accumulation of dissipation. Unlike previous nonreciprocal studies~\cite{haldane2008possible,zhang2021superior,sounas2013giant,popa2014non,nassar2020nonreciprocity}, neither magnetic fields nor spatiotemporal modulation is required. Instead, our magnetic-free nonreciprocity is achieved by dynamically encircling a nexus~\cite{zhou2023higher}, which benefits from the robustness provided by bistability.
Moreover, these nonreciprocal processes provide a new method to design nonreciprocal entanglement, different from methods that rely on dynamically encircling an exceptional point~\cite{li2023speeding} or the Sagnac effect~\cite{jiao2020nonreciprocal}. Utilizing this method, we demonstrate, for the first time, nonreciprocal entanglement in nonlinear systems, characterized by robustness and stability. More broadly, this method provides a general approach for designing nonreciprocal transmission and entanglement in various nonlinear physical systems~\cite{boyd2008nonlinear,wang2023image,yildirim2024nonlinear,zhang2023graphene,pisarchik2014control,bachtold2022mesoscopic,peyronel2012quantum,chang2014quantum,roberts2020driven,yesharim2023direct,wang2023quantum,gao2023quantum,yang2021bistability,yang2023nonlinearity,norte2016mechanical}.

\begin{figure}[ht]
    \centering \includegraphics[width=1.0
    \linewidth]{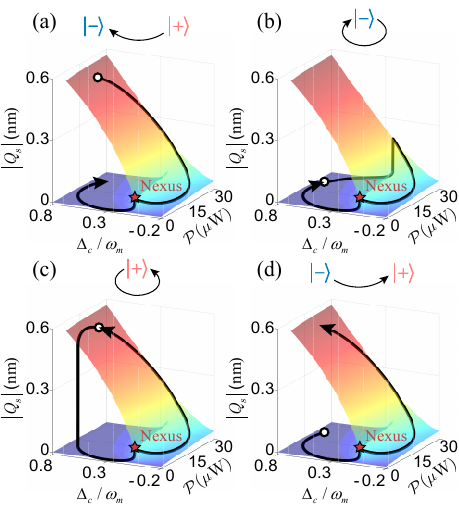}
    \caption{\textbf{Nonreciprocal displacements induced by bistability:} Dynamically encircling the nexus clockwise from (a) the upper branch $|+\rangle$ and (b) the lower branch $|-\rangle$ results in $|-\rangle$. Dynamically encircling the nexus counterclockwise from (c) the upper branch $|+\rangle$ and (d) the lower branch $|-\rangle$ leads to $|+\rangle$.}
    \label{fig2}
\end{figure}

\begin{figure}[ht]
    \centering \includegraphics[width=1.0
    \linewidth]{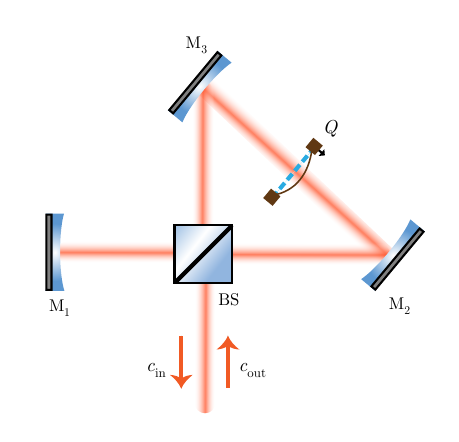}
    \caption{\textbf{Schematic of the dissipatively coupled optomechanical system.} $\rm{M}_1$, $\rm{M}_2$, and $\rm{M}_3$ are perfectly fixed mirrors, while $\rm{BS}$ denotes a fixed beamsplitter.}
    \label{fig1}
\end{figure}

{\it Physical mechanism and model:} As illustrated in Fig.~\ref{fig2}, a typical surface of mechanical displacement $|Q|$, arising from nonlinearity, is plotted with two control parameters: detuning $\Delta_c$ and input power ${\cal P}$. 
The nonlinear surfaces reveal regions of monostability and bistability, separated by transition points called bifurcation points. At the "pitchfork" bifurcation point, these bistable surfaces converge to form a cusp known as the nexus~\cite{zhou2023higher}. Using this special structure, we design trajectories encircling the nexus in the parameter space and set start points in the bistable regime. In Figs.~\ref{fig2}(a) and (b), we set the starting points of the trajectories located on the upper branch $|+\rangle$ and lower branch $|-\rangle$ of the bistability surfaces, respectively. The clockwise evolutions drive both $|+\rangle$ and $|-\rangle$ to $|-\rangle$, whereas the counterclockwise evolutions steer both $|+\rangle$ and $|-\rangle$ to $|-\rangle$. 
These direction-dependent outcomes of the mechanical displacement $|Q_s|$ exhibit nonreciprocal behaviors by dynamically encircling a nexus in the parameter space.

As shown in Fig.~\ref{fig1}, to illustrate such nonreciprocal behaviors, we consider a dissipative and dispersive optomechanical system~\cite{sawadsky2015observation}. It consists of a mechanical resonator (frequency $\omega_m$ and decay rate $\gamma$) and a cavity mode ${c}$ (frequency $\omega_c$ and decay rate $\kappa$). 
An external field drives the cavity mode ${c}$, which dispersively and dissipatively couples the mechanical mode ${b}$ with strengths $g_\omega$ and $g_\kappa$, respectively. 
In a frame rotating at the drive frequency $\omega _d$, our system is described by the Hamiltonian
$
     H = \hbar {\Delta _c}(Q){c^\dag }c + \frac{1}{2}(m\omega _m^2{Q^2} + \frac{{{P^2}}}{m}) + i\hbar \sqrt {\kappa (Q)} ({c_{in}}{c^\dag } - {c_{in}}^\dag c),
$
where the effective detuning is ${\Delta _c}\left( Q \right) = {\Delta _c} + {g_\omega } Q $ with ${\Delta _c} = {\omega _c} - {\omega _d}$, and the effective decay is $\kappa \left(  Q \right) = \kappa (1 + {g_\kappa } Q/\kappa )$, $Q$ and $P$ denote the position and momentum of the mechanical resonator with mass $m$ and frequency $\omega_m$, respectively. The dispersive and dissipative coupling strengths are defined as ${g_\omega } = \frac{{d{\omega _c}(Q)}}{{dQ}}$ and ${g_\kappa } = \frac{{d\kappa (Q)}}{{dQ}}$, respectively. The drive strength ${\varepsilon }{ = }\sqrt {{\cal P}/\left( {\hbar {\omega _d}} \right)}$ is determined by the input power ${\cal P}$ and frequency $\omega _d$.

To fully explore the nonlinear dynamics of the system, we account for cavity dissipation and environmental noise, resulting in the following Langevin equations of motion:
\begin{equation}
\begin{aligned}
\dot{c} &= - \left[ \frac{\kappa + g_\kappa Q}{2} + i(\Delta_c + g_\omega Q) \right] c + \sqrt{\kappa + g_\kappa Q} \, c_{in} \\
\dot{Q} &= \frac{1}{m} P \\
\dot{P} &= - m \omega_m^2 Q - \hbar g_\omega c^\dag c - i \frac{\hbar g_\kappa}{2\sqrt{\kappa}} (c_{in} c^\dag - c_{in}^\dag c) 
           \\
           &- \gamma P +  \xi
\end{aligned}
\label{Langevin}
\end{equation}
where the mechanical noise operator $\xi$ satisfies $\left\langle {\xi (t)\xi (t')} \right\rangle  = \hbar m{\omega _m}\gamma (2{\left[ {\exp \left( {\hbar {\omega _m}/{k_B}\cal{T}} \right) - 1} \right]^{ - 1}} + 1)\delta (t - t')$ with the temperature $\cal{T}$~\cite{xuereb2011dissipative}. The input field $c_{in}=\varepsilon +{ a}_{in}$ consists of a drive $\varepsilon $ and vacuum input noise ${ a}_{in}$, which satisfies the correlation function$\left\langle {{{ a}_{in}}\left( t \right){{ a}^{\dag}_{in} }\left( {t'} \right)} \right\rangle  = \delta \left( {t - t'} \right)$. In the above Langevin Eq.~(\ref{Langevin}), the commutation relation $[P,\sqrt {\kappa (Q)} ] =  - i\frac{{{g_\kappa }}}{{2\sqrt \kappa  }}$ is obtained by approximation $\sqrt {\kappa  + {g_\kappa }Q}  \approx \sqrt \kappa   + \frac{{{g_\kappa }Q}}{{2\sqrt \kappa  }}$ with ${g_\kappa }Q \ll \kappa$.

To characterize the quantum properties of the system, we express each operator as the sum of its steady-state value and quantum fluctuation: ${Q}=Q_s+q$, ${P}=P_s+p$, ${c}=c_s+a$ and ${c_{in}} = \varepsilon  + a_{in}$. The steady-state values are given by $ P_s  = 0$, ${Q_s} =  - \frac{{\hbar {g_\omega }}}{{m\omega _m^2}}|{c_s}{|^2} - i\frac{{\hbar {g_\kappa }\varepsilon }}{{2m\omega _m^2\sqrt \kappa  }}({c_s}^ *  - {c_s})$, and ${c_s} = \sqrt {{\kappa _{eff}}} \varepsilon /(\frac{{\kappa _{eff}}}{2} + i{\Delta_{eff}})$ with effective detuning ${\Delta_{eff}} = {\Delta _c} + {g_\omega }Q_s$ and effective decay ${\kappa _{eff}} = \kappa  + {g_\kappa }Q_s$.
After neglecting the small terms, the linearized quantum Langevin equations for the fluctuation operators take the form:
\begin{equation}
\begin{aligned}
\dot a &= - \left( \frac{\kappa_{\text{eff}}}{2} + i \Delta_{\text{eff}} \right) a 
         - \left( \frac{G_\kappa}{2} + i G_\omega - \frac{g_\kappa \varepsilon}{2\sqrt{\kappa_{\text{eff}}}} \right) q \\
         &+ \sqrt{\kappa_{\text{eff}}} a_{\text{in}} \\
\dot q &= \frac{1}{m} p \\
\dot p &= - m \omega_m^2 q 
         - \hbar \left( G_\omega^* a + G_\omega a^\dag \right) 
         - i \frac{\hbar g_\kappa \varepsilon}{2 \sqrt{\kappa}} \left(  a^\dag -  a \right) \\
       &\quad - i \frac{\hbar}{2 \sqrt{\kappa}} \left( G_\omega^* a_{\text{in}} - G_\omega a_{\text{in}}^\dag \right) 
         - \gamma p 
         +  \xi
\end{aligned}
\label{quantum}
\end{equation}
where $G_\omega=g_\omega c_s$ and $G_\kappa=g_\kappa c_s$ represent the linear coupling strengths for $g_\omega$ and $g_\kappa$, respectively.

{\it Nonreciprocal behaviors of displacements:} The steady values above show that the nonlinear effective decay rate and radiational pressure coupling could result in a bistable surface for $|Q_s|$ with different $\Delta_c$ and ${\cal P}$ as plotted in Fig.~\ref{fig2}. Here, we set the experimentally achievable parameters: the mechanical mode quality factor ${\cal Q}=5.8\times10^{5}$, the environmental temperature ${\cal{T}}=0.5\rm{mK}$, the effective mass of the membrane $m = 80{\rm{ng}}$, the mechanical mode frequency ${\omega _m}  = 136{\rm{kHz}}$,  the cavity decay rate $\kappa=0.1\omega _m$, the laser wavelength $\lambda=1064\rm{nm}$, the dispersive coupling strength  ${g_\omega } = 196.57{\rm{kHz/nm}}$, and the dissipative coupling strength ${g_\kappa } = 17.47{\rm{kHz/nm}}$.
Thus, dynamically encircling the nexus point in the parameter space can be achieved through parameter modulation: ${\cal P} = {\cal P}_{0} + A\cos (\theta  + {\theta _0})$ and ${\Delta _c} = {\Delta _{0}} + B\sin (\theta  + {\theta _0})$, where $A=A_0(1+\delta)$ and $B=B_0(1+\delta)$ define the trajectory with fluctuation $\delta$, $\theta$ is the encircling angle, and $\theta _0$, the initial conditions of the trajectory follow ${\theta _0}=0.28\pi$, ${\cal P} _{0}=15{\mu W}\ $, ${\Delta _{0}}=0.3\omega_{m}$, $B_0=0.45 \omega _{m}$ and $A_0=15 \rm{\mu W}$. In addition, we introduce the fluctuation $\delta$ to demonstrate the robustness of our nonreciprocal behaviors and set ${\cal P}\geq0$ in the following simulations.

\begin{figure}[hb]
\centering
\vspace{0pt} 
\includegraphics[width=1.0\linewidth]{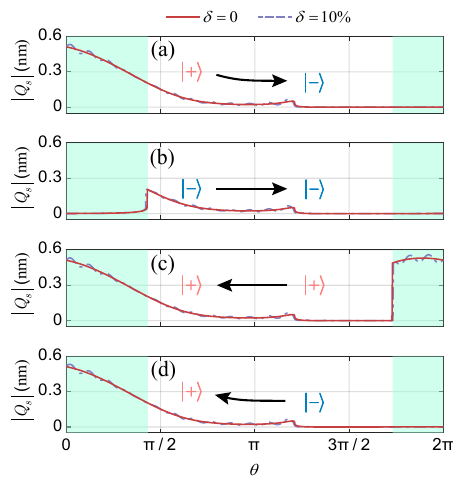}
\begin{minipage}[h]{0.5\textwidth}
\caption{\textbf{Nonreciprocal evolution of the displacement $|Q_s|$ under dynamic parameter modulation of $\theta$.} \textbf{(a)} Clockwise evolution from the upper branch $|+\rangle$ to the lower branch $|-\rangle$. \textbf{(b)} Clockwise Evolution from the lower branch $|-\rangle$ to the lower branch $|-\rangle$. \textbf{(c)} Counterclockwise Evolution returning from the upper branch $|+\rangle$ to  the upper branch $|+\rangle$. \textbf{(d)} Counterclockwise evolution from the lower branch $|-\rangle$ to the upper branch $|+\rangle$. The red solid lines depict the idea evolution along the trajectories. The dash-dotted purple curves represent the evolution with a 10\% fluctuation. Here the green shaded area represents the bistable region.}
\label{fig3}
  \end{minipage}
\end{figure}

Figure~\ref{fig3} displays the nonreciprocal behavior of displacement $|Q_s|$ from Fig.~\ref{fig1}, plotted as a function of $\theta$, achieved by dynamically encircling a nexus with the starting point in the bistable regime. Figs.~\ref{fig3}(a) and (b) exhibit the evolution in the clockwise encircling the nexus. In Fig.~\ref{fig3}(a), the initial state is $|+\rangle$ on the upper branch of the bistability. As $\theta$ increases from $0$ to $2\pi$, we observe that the encircling evolution follows the upper branch of the bistability regime, evolves to the monostability regime without transitions, and returns to the bistability regime, resulting in the final state $|-\rangle$ on the lower branch of the bistability. In Fig.~\ref{fig3}(b), the initial state is $|-\rangle$. Increasing $\theta$ from $0$ to $2\pi$, we find that the evolution transitions from the lower branch in the bistable regime to the monostable regime and then returns to the initial state $|-\rangle$. In contrast, Figs.~\ref{fig3}(c) and (d) show counterclockwise evolutions with initial states $|+\rangle$ and $|-\rangle$, respectively. For these counterclockwise evolutions, dynamically encircling the nexus leads to a final state $|+\rangle$.

These dynamical evolutions demonstrate that the final states depend only on the direction of encircling the nexus, independent of the initial states. These nonreciprocal behaviors are ensured by the stability of the bistability and monostability when the starting points are set in the bistable regime. Therefore, for a closed evolution trajectory, our simulations confirm that the final states are determined only by the evolution directions and the parameters of the starting points, exhibiting robustness against parameter fluctuations with deviations as large as
10\% (see the purple curves in Fig.~\ref{fig3}).

These nonreciprocal behaviors are strongly reminiscent of those in non-Hermitian systems for the exceptional point~\cite{doppler2016dynamically,zhang2018dynamically,liu2021dynamically,zhang2019dynamically,liu2020efficient,li2020hamiltonian,nasari2022observation,chen2022decoherence}. Nevertheless, they originate from distinct physical mechanisms and differ fundamentally in stability. In the non-Hermitian system, nonreciprocal behaviors arise from a combination of stable and unstable states, resulting in their vanishing for a long evolution time~\cite{hassan2017dynamically,gao2025photonic,bu2024chiral,zhang2022topological}. Conversely, by dynamically encircling a nexus of the nonlinear system, our nonreciprocal behaviors are independent of evolution time and protected by stability.

\begin{figure*}[ht]
    \centering \includegraphics[width=18cm]{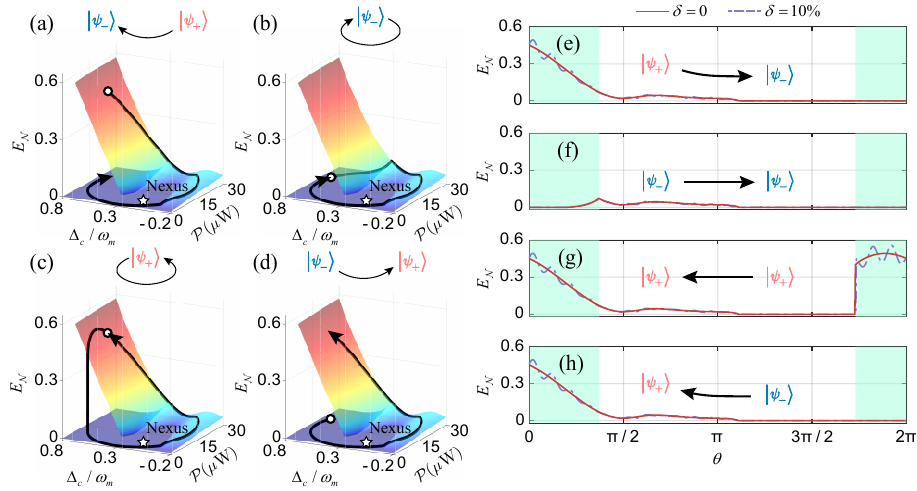}
    \caption{\textbf{Nonreciprocal entanglement by encircling the nexus.} \textbf{(a)} Clockwise evolution from the upper branch (entangled state $|\psi_{+}\rangle$) to the lower branch (the thermal state $\psi_{-}$).~\textbf{b)} Clockwise evolution returning from the lower branch ($|\psi_{-}\rangle$) to the lower branch ($\psi_{-}\rangle$).~\textbf{(c)} Counterclockwise evolution retuning from the upper branch ($|\psi_{+}\rangle$) to the upper branch ($|\psi_{+}\rangle$). \textbf{(d)} Counterclockwise evolution from the lower branch ($\psi_{-}$) to the upper branch ($\psi_{+}$). The corresponding dynamic parameter modulations of $\theta$ are presented in \textbf{(e)-(f)}. The red solid lines represent the idea trajectories without perturbation ($\delta = 0$). The dash-dotted purple lines indicate trajectories with a $\delta = 10\%$ fluctuation, exhibiting robustness. The shaded area denotes the bistable region.}
    \label{fig4}
\end{figure*}

{\it Nonreciprocal entanglements:} Such nonreciprocal behavior holds a potential for designing nonreciprocal entanglement. To exhibit this physical phenomenon, we define optical quadrature operators $X = (a + {a^\dag })/\sqrt 2 $ and $Y = (a - {a^\dag })/(\sqrt 2 i)$, and their noises $X_{\rm{in}}=({a}_{\rm{in}} + a_{\rm{in}}^\dag )/\sqrt {2}$ and $Y_{\rm{in}}= (a_{in} - a_{in}^\dag)/(\sqrt{2} i)$, respectively. We then rewrite Eq.~(\ref{quantum}) as ~\cite{vitali2007optomechanical}
\begin{equation}\label{u}
\dot u(t) = Au(t) + n(t)
\end{equation} with ${u^T} = ( X, Y, q, p)$ (the superscript $T$ denotes the transposition),
\begin{equation}
\resizebox{\linewidth}{!}{$
A = \left(
\begin{array}{cccc}
- \frac{\kappa_{\text{eff}}}{2} & \Delta_{\text{eff}} &
{ - \frac{1}{{\sqrt 2 }}{\mathop{\rm Re}\nolimits} ({G_\kappa }) + \sqrt 2 {\mathop{\rm Im}\nolimits} ({G_\omega }) + \frac{{{g_\kappa }\varepsilon }}{{\sqrt {2{\kappa _{eff}}} }}} & 0 \\
- \Delta_{\text{eff}} & - \frac{\kappa_{\text{eff}}}{2} &
{ - \frac{1}{{\sqrt 2 }}{\mathop{\rm Im}\nolimits} ({G_\kappa }) - \sqrt 2 {\mathop{\rm Re}\nolimits} ({G_\omega })} & 0 \\
0 & 0 & 0 & \frac{1}{m} \\
{ - \hbar \sqrt 2 {\mathop{\rm Re}\nolimits} ({G_\omega })} & { - \hbar \sqrt 2 {\mathop{\rm Im}\nolimits} ({G_\omega }) - \frac{{\hbar {g_\kappa }\varepsilon }}{{\sqrt {2\kappa } }}} &
- m \omega_m^2 & - \gamma
\end{array}
\right),
$}
\label{compactA}
\end{equation} 
and ${n^T} = (\sqrt{\kappa _{eff}} {X_{{\rm{in}}}}, \sqrt{{\kappa _{eff}} } {Y_{{\rm{in}}}}, 0, - \frac{{\hbar }}{{\sqrt {2\kappa } }}{\mathop{\rm Im}\nolimits} ({G_\kappa}){X_{in}} + \frac{{\hbar }}{{\sqrt {2\kappa } }}{\mathop{\rm Re}\nolimits} (G_\kappa){Y_{in}} + \xi )$.

Under the stability condition, we gets the steady-state correlation matrix~\cite{vitali2007optomechanical}:
\begin{equation}
\begin{split}
AV + V{A^T} = -D,
\end{split}
\label{Lya}
\end{equation}
which enables us to determine $V$ for any given parameter values in the evolution trajectories, where $D = {\rm{diag}}[{\frac{{{\kappa _{\rm{eff}}}}}{2}}, {\frac{{{\kappa _{\rm{eff}}}}}{2}},0,\hbar m{\omega _m}\gamma \left\{2{\left[ {\exp \left( {\hbar {\omega _m}/{k_B}\cal{T}} \right) - 1} \right]^{ - 1}} + 1\right\} + \frac{{{\hbar ^2}G_\kappa ^2}}{{4\kappa }}]$ is the diffusion matrix stemming from the noise correlations.

To quantify entanglement between the mechanical resonator and the optical mode, we take logarithmic negativity ${E_{\cal N}}= \max [0, - \ln (2{\eta ^ - })]$ as a measure of entanglement for continuous variables~\cite{vidal2002computable}. Here $\eta^-=2^{-1/2}\{\sum(V)-[\sum(V)^2-4\mathrm{det}V]^{1/2}\}^{1/2}$, $\sum(V)=\mathrm{det} V_A+\mathrm{det} V_B-2\mathrm{det} V_c$, and we utilize the $2\times2$ block form of the correlation matrix:
\begin{equation}
V=\left(
\begin{array}{cccc}
V_A&V_C\\
V_C&V_B\end{array}
\right)
\end{equation}
Thus, $\eta^-<1/2$ corresponds to Simon’s necessary and sufficient criterion for entanglement via non-positive partial transpose for Gaussian states~\cite{simon2000peres,vitali2007optomechanical}.

In this manner, we can quantify the steady-state entanglement of the dissipatively coupled optomechanical system using ${E_{\cal N}}$. 
Eq.~(\ref{u}) reveals that steady-state entanglement depends directly on semi-classical mean physical quantities, such as displacement $|Q_s|$, enabling control of quantum entanglement by adjusting these quantities. In Fig.~\ref{fig4}(a-d), we numerically calculate the steady-state entanglements, which exhibit behavior and trends similar to $|Q_s|$ in Fig.~\ref{fig1}. The upper and lower branches of the bistability surfaces correspond to the entanglement $|\psi _{+}\rangle$ and the thermal states $|\psi _{-}\rangle$, respectively. This not only suggests quantum state transfer between these states but also indicates that quantum states exhibit nonreciprocal behavior by dynamic encircling of the nexus, as shown in Fig.~\ref{fig4}(e-f). 

In Fig.~\ref{fig4}, we define trajectories identical to those in Fig.~\ref{fig3}, with evolutions starting from the bistable regime (highlighted in green). Figs.~\ref{fig4} (a,e) and (b,f) show clockwise evolutions with initial states $|\psi _{+}\rangle$ and $|\psi _{-}\rangle$, respectively. Counterclockwise evolutions for the same initial states $|\psi _{+}\rangle$ and $|\psi _{-}\rangle$ are depicted in Figs.~\ref{fig4} (c,g) and (d,h), respectively. Therefore, one can see that clockwise evolutions result in the final thermal states $|\psi _{-}\rangle$ with $E_{\cal N}=0$, while counterclockwise evolutions only yield entangled states $|\psi_{+}\rangle$ with $E_{\cal N}\approx0.46$. We note that our nonreciprocal behavior of quantum states features stability and robustness against parameter fluctuations in closed trajectories with the highest value of $E_{\cal N}\approx0.50$ [see Figs.~\ref{fig4} (g)]. These properties
make our nonreciprocal entanglements more stable than those achieved by dynamically encircling an exceptional point~\cite{hassan2017dynamically,gao2025photonic,bu2024chiral,zhang2022topological}, where nonreciprocity might be lost over long-term evolutions.

Our simulations reveal that high entanglement between mechanical and optical modes corresponds to a large displacement $|Q_s|$. This arises from that a large $|Q_s|$ indicates a large mean photon number $\langle n\rangle=\langle c_s^{\dag}c_s\rangle$, which enhances the dispersive and dissipative coupling strengths ($|G_{\omega}|=|g_{\omega}\sqrt{n}|$ and $|G_{\kappa}|=|g_{\kappa}\sqrt{n}|$) and decreases the effective dissipation ($\kappa_{eff}=\kappa-g_{\kappa}|Q_s|$).
Specifically, the lower branch of $|Q_s|$ corresponds to high effective dissipation ($\kappa_{eff} \approx \kappa$) with weak dispersive and dissipative coupling strength, while the upper branch exhibits low effective dissipation ($\kappa_{eff} \approx 0.5\kappa$) with a strong dispersive and dissipative coupling strength. 
These parameters suggest that strong coupling and low effective dissipation can effectively generate entanglement between the mechanical and optical modes, whereas weak coupling and high effective dissipation lead to a thermal state, as the entanglement is masked by thermal noise. Besides, the external drive can inject coherence into the optomechanical system, ensuring that steady-state entanglement remains robust against decoherence. By comparing Figs.~\ref{fig3} and~\ref{fig4}, we observe that the upper branch of $|Q|$ corresponds to an entangled state with a high $E_{\cal N}$, while the lower branch yields a thermal state with $E_{\cal N}=0$.

More importantly, we highlight the nonreciprocal behavior of quantum states, achieved by dynamically encircling a nexus of bistability. Our physical mechanism offers a practical approach for designing nonreciprocal quantum devices, leveraging the stability of the bistability surfaces. It is reminiscent of other forms of quantum nonreciprocity, such as those arising from dynamically encircling an exceptional point~\cite{li2023speeding}, quantum squeezing ~\cite{tang2022quantum}, the Sagnac effect~\cite{jiao2020nonreciprocal,bin2024nonreciprocal}, or symmetry breaking induced by engineered reservoirs~\cite{orr2023high}.

{\it Conclusion:}  In summary, we have presented how to design quantum nonreciprocal entanglement and adjust nonreciprocal behaviors, spanning classical to quantum regimes, by dynamically encircling a nexus in a dissipative coupled optomechanical system. These nonreciprocal behaviors are characterized by stability and robustness against variations in closed trajectories arising from the intrinsic stability of nonlinear physics. Moreover, our work advances the development of quantum nonreciprocity, progressing from the first-order linear nonreciprocity~\cite{li2023speeding}, to second-order squeezing nonreciprocity~\cite{tang2022quantum}, and finally to our three-order bistable nonreciprocity. Finally, our study, integrating nonlinear physics with quantum engineering, opens a new avenue for designing nonreciprocal and robust quantum devices in nonlinear systems and provides opportunities to explore nonreciprocal quantum properties through nonlinear effects in optics~\cite{abraham1982optical}, atomic gases~\cite{ding2020phase}, electronics~\cite{kurpiers2018deterministic}, and acoustics~\cite{chen2020bistability}.

\section*{Acknowledgments}
This work was supported by National Natural Science Foundation of China under Grant Nos. 12564047, 11874004, 11204019, 12564048, 12204137.

\bibliography{article08204}
\end{document}


\preprint{APS/123-QED}

\title{\textbf{Supplementary Materials:\\
Nonreciprocal Entanglement by Dynamically Encircling a Nexus} 
}%

\author{Lei Huang}
 \affiliation{College of Physics and Electronic Engineering, Hainan Normal University, Haikou 571158, China.}
 
 \author{Peng-Fei Wang}
 \affiliation{College of Physics and Electronic Engineering, Hainan Normal University, Haikou 571158, China.}
 
 \author{Jian-Qi Zhang}
 \email{changjianqi@gmail.com}
 \affiliation{State Key Laboratory of Magnetic Resonance and Atomic and Molecular Physics, Wuhan Institute of Physics and Mathematics, Chinese Academy of Sciences, Wuhan 430071, China}
 
 \author{Xin Zhou}
 \affiliation{College of Intelligence Science and Technology, National University of Defense Technology, Changsha 410073, China.}

  \author{Shuo Zhang}
 \affiliation{Henan Key Laboratory of Quantum Information and Cryptography, Zhengzhou 450000, China}
 
 \author{Hong Yang}
 \affiliation{College of Physics and Electronic Engineering, Hainan Normal University, Haikou 571158, China.}

 \author{Dong Yan}
 \affiliation{College of Physics and Electronic Engineering, Hainan Normal University, Haikou 571158, China.}

\date{\today}

\maketitle


\section{\label{Derivation of the Hamiltonian}Theoretical Model and System Hamiltonian}

We consider a hybrid optomechanical system that features both dispersive and dissipative couplings. Its total Hamiltonian is given by~\cite{elste2009quantum,xuereb2011dissipative}

\begin{equation}
H = \hbar {\omega _c}{c^\dag }c + \frac{1}{2}(m\omega _m^2{Q^2} + \frac{{{P^2}}}{m}) + {H_\kappa } + {H_{{\mathop{\rm int}} }} + {H_\gamma },
\label{H}
\end{equation}
where  $H_\gamma$ describes the intrinsic mechanical damping by an equilibrium bath at temperature ${\cal T}$. ${c^\dag }$ and $c$ are the creation and annihilation operators of the cavity field with frequency ${\omega _c}$, and $Q$, $P$ represent the position and momentum operators of the mechanical resonator with mass m and frequency ${\omega _m}$, respectively. The cavity-bath interaction takes the form of

\begin{equation}
{H_\kappa } = i\hbar \sqrt \kappa  \int {\frac{{d\omega }}{{\sqrt {2\pi } }}({c_\omega }{c^\dag } - H.c.)}.
\label{Hkappa}
\end{equation}

The cavity-mechanical coupling is described by

\begin{equation}
{H_{{\mathop{\rm int}} }} = \hbar {g_\omega }Q{c^\dag }c + i\hbar \frac{{{g_\kappa }}}{{2\sqrt \kappa  }}Q\int {\frac{{d\omega }}{{\sqrt {2\pi } }}(c_\omega ^\dag c - H.c.)},
\label{Hint}
\end{equation}
where $g_{\omega}$ and $g_{\kappa}$ are the dispersive and dissipative optomechanical coupling strengths, respectively. The first term represents the standard radiation-pressure (dispersive) interaction, whereas the second arises from the position-dependent cavity linewidth, giving rise to dissipative coupling. The operator $c_{\omega}$ represents the external field mode at frequency $\omega$.

To derive the Heisenberg equations of motion, we adapt the input–output formalism to dissipative coupling. This leads to the following expression:~\cite{weiss2013strong}

\begin{equation}
\sqrt \kappa  \int {\frac{{d\omega }}{{\sqrt {2\pi } }}{c_\omega }}  = \sqrt \kappa  {c_{\rm{in}}} + (\frac{\kappa }{2} + \frac{{{g_\kappa }Q}}{4})c,
\label{in-out}
\end{equation}
where $c_{\rm{in}}$ is the optical input mode. The input–output relation is given by

\begin{equation}
{c_{{\rm{in}}}} - {c_{{\rm{out}}}} =  - \sqrt \kappa  c - \frac{{{g_\kappa }}}{{2\sqrt \kappa  }}Qc.
\label{input-output}
\end{equation}

Using Eq.~(\ref{in-out}), we then obtain the Heisenberg–Langevin equations in the rotating frame, which describe the full nonlinear dynamics of the coupled cavity and mechanical modes:

\begin{widetext}
\begin{equation}
\begin{aligned}
{\dot c} &=  - i(\Delta  + {g_\omega }Q)c - (\frac{\kappa }{2} + \frac{{{g_\kappa }Q}}{2})c + \sqrt {\kappa  + {g_\kappa }Q} {c_{{\rm{in}}}}, \\
{\dot Q} &= \frac{1}{m}P,\\
{\dot P} &=  - m\omega _m^2Q - \hbar {g_\omega }{c^\dag }c - i\frac{{\hbar {g_\kappa }}}{{2\sqrt \kappa  }}({c_{{\rm{in}}}}{c^\dag } - c_{{\rm{in}}}^\dag c) - \gamma P + \xi.
\end{aligned}
\label{langvin}
\end{equation}
\end{widetext}

\section{\label{Optomechanical Bistability and Cubic Steady-State Solutions}Optomechanical Bistability and Cubic Steady-State Solutions}

Each Heisenberg operator can be decomposed into a steady-state component and a fluctuation term as follows: $Q = Q_s + q$, $P = P_s + p$, $c = c_s + a$, and $c_{\rm{in}} = \varepsilon + a_{\rm{in}}$, where the drive strength ${\varepsilon }{ = }\sqrt {{\cal P}/\left( {\hbar {\omega _d}} \right)}$ is determined by the input power ${\cal P}$ and frequency $\omega _d$. The steady-state solutions are obtained from Eq.~(\ref{langvin}) as

\begin{widetext}
\begin{equation}
\begin{aligned}
P_s  &= 0,\\
{Q_s} &=  - \frac{{\hbar {g_\omega }}}{{m\omega _m^2}}|{c_s}{|^2} - i\frac{{\hbar {g_\kappa }\varepsilon }}{{2m\omega _m^2\sqrt \kappa  }}({c_s}^ *  - {c_s}),\\
{c_s} &= \sqrt {{\kappa _{eff}}} \varepsilon /(\frac{{\kappa _{eff}}}{2} + i{\Delta_{eff}}).\\
\end{aligned}
\label{stable}
\end{equation}
\end{widetext}
with effective decay rate ${\kappa_{\mathrm{eff}}} = \kappa + g_{\kappa} Q_s$ and effective detuning ${\Delta_{\mathrm{eff}}} = \Delta_c + g_{\omega} Q_s$.

In particular, the steady-state displacement $Q_s$ of the mechanical resonator satisfies the following cubic nonlinear equation

\begin{equation}
{d_1}Q_s^3 + {d_2}Q_s^2 + {d_3}{Q_s} + {d_4} = 0,
\label{3th}
\end{equation}
with
\begin{widetext}
\begin{equation}
\begin{aligned}
{d_1} &= m\omega _m^2(\frac{{g_\kappa ^2}}{4} + g_\omega ^2),
\\
{d_2} &= m\omega _m^2(\frac{{\kappa {g_\kappa }}}{2} + 2\Delta {g_\omega }) - \frac{{\hbar g_\kappa ^2{g_\omega }{\varepsilon ^2}}}{{2\kappa }},
\\
{d_3} &= m\omega _m^2(\frac{{{\kappa ^2}}}{4} + {\Delta ^2}) - \frac{{\hbar g_\kappa ^2\Delta {\varepsilon ^2}}}{{2\kappa }},
\\
{d_4} &= - \hbar {\varepsilon ^2}({g_\kappa }\Delta  - {g_\omega }\kappa ).
\end{aligned}
\label{ddd}
\end{equation}
\end{widetext}
This cubic equation permits the emergence of bistable behavior in our dissipative optomechanics. These solutions can be obtained analytically using the standard formula for cubic equations. The existence of triple real roots corresponds to the emergence of optical bistability in the system.

In the bistable region, the three steady-state solutions of mechanical displacement are given by
\begin{equation}
{Q_k} = 2\sqrt { - \frac{{3B - {A^2}}}{9}} \cos \left( {\frac{{\theta  + 2k\pi }}{3}} \right) - A/3,k = 0,1,2,
\label{yk}
\end{equation}
where $\theta = \arccos \left( \frac{2A^3 - 3AB + 9C}{2(3B - A^2)\sqrt{A^2 - 3B}} \right)$ and coefficients are defined as $A = \frac{{{d_2}}}{{{d_1}}}$, $B = \frac{{{d_3}}}{{{d_1}}}$ and $ C = \frac{{{d_4}}}{{{d_1}}}$.

\section{\label{Stability conditions}Linearized Dynamics and Stability Conditions}

Based on the previously derived steady-state solution, we obtain the linearized quantum Langevin equations as

\begin{widetext}
\begin{equation}
\begin{aligned}
\dot a &=  - i{g_\omega }{c_s}q - i{\Delta _{{\rm{eff}}}}a - \frac{{{\kappa _{{\rm{eff}}}}}}{2}a - \frac{{{g_\kappa }}}{2}{c_s}q + \frac{{{g_\kappa }\varepsilon }}{{2\sqrt \kappa  }}q - \sqrt {{\kappa _{{\rm{eff}}}}} {{\hat a}_{\rm{in}}},\\
\dot q &= \frac{1}{m}p,\\
\dot p &=  - m\omega _m^2q - \hbar {g_\omega }c_s^ * a + {c_s}{a^\dag } - i\frac{{\hbar {g_\kappa }}}{{2\sqrt \kappa  }}(c_s^ * {a_{{\rm{in}}}} - {c_s}a_{{\rm{in}}}^\dag ) - i\frac{{\hbar {g_\kappa }\varepsilon }}{{2\sqrt \kappa  }}({a^\dag } - a) - \gamma \hat p + \xi .
\end{aligned}
\label{quantum}
\end{equation}
\end{widetext}

The cavity field fluctuation quadratures are defined as $X = (a + {a^\dag })/\sqrt 2 $ and $Y = (a - {a^\dag })/\sqrt 2 i$, along with the input noise operators ${X_{{\rm{in}}}} = ({a_{{\rm{in}}}} + a_{{\rm{in}}}^\dag )/\sqrt 2 $ and ${Y_{{\rm{in}}}} = ({a_{{\rm{in}}}} - a_{{\rm{in}}}^\dag )/\sqrt 2 i$. With these definitions, the linearized quantum Langevin equations (QLEs) can be expressed as

\begin{equation}
\dot u(t) = Au(t) + n(t),
\label{compact}
\end{equation}
where $u^T(t)={\left( X\left( t \right),Y\left( t \right), {q\left( t \right),p\left( t \right)} \right)^T}$ (the superscript $T$ denotes the transposition) is vector of continuous-variable fluctuation operators, and the corresponding vector of noises is
\begin{equation}
{n^T}\left( t \right) = \left( {\sqrt {{\kappa _{eff}}} {X_{in}}\left( t \right),\sqrt {{\kappa _{eff}}} {Y_{in}}\left( t \right),0, - \frac{{\hbar {g_\kappa }}}{{\sqrt {2\kappa } }}{\mathop{\rm Im}\nolimits} ({c_s}){X_{in}}\left( t \right) + \frac{{\hbar {g_\kappa }}}{{\sqrt {2\kappa } }}{\mathop{\rm Re}\nolimits} ({c_s}){Y_{in}}\left( t \right) + \frac{\hbar }{{{Q_{zpf}}}}\xi \left( t \right)} \right),
\label{vector of noises}
\end{equation}
the matrix $A$ is in the form of
\begin{equation}
A = \left( {\begin{array}{*{20}{c}}
{ - \frac{{{\kappa _{eff}}}}{2}}&{{\Delta _{eff}}}&{{u_1}}&0\\
{ - {\Delta _{eff}}}&{ - \frac{{{\kappa _{eff}}}}{2}}&{{u_2}}&0\\
0&0&0&{\frac{1}{m}}\\
{{v_1}}&{{v_2}}&{ - m\omega _m^2}&{ - \gamma }
\end{array}} \right),
\label{A}
\end{equation}
with ${u_1} =  - \frac{{{g_\kappa }}}{{\sqrt 2 }}{\mathop{\rm Re}\nolimits} ({c_s}) + \frac{{{g_\kappa }\varepsilon }}{{\sqrt {2{\kappa _{eff}}} }} + \sqrt 2 {g_\omega }{\mathop{\rm Im}\nolimits} ({c_s})$, ${u_2} =  - \frac{{{g_\kappa }}}{{\sqrt 2 }}{\mathop{\rm Im}\nolimits} ({c_s}) - \sqrt 2 {g_\omega }{\mathop{\rm Re}\nolimits} ({c_s})$, ${v_1} =  - \hbar {g_\omega }\sqrt 2 {\mathop{\rm Re}\nolimits} ({c_s})$ and ${v_2} =  - \hbar {g_\omega }\sqrt 2 {\mathop{\rm Im}\nolimits} ({c_s}) - \frac{{\hbar {g_\kappa }\varepsilon }}{{\sqrt {2\kappa } }}$. The coefficients ${u_i}$ and ${v_i}$ denote the effective radiation-pressure and dissipative back-action contributions to the fluctuation dynamics. 
According to the Routh–Hurwitz criterion, the system is stable if the following conditions are satisfied:~\cite{gradshteyn2014table}:

\begin{widetext}
\begin{equation}
\begin{aligned}
s_1 &= \omega_m^2\left(\frac{\kappa_{\text{eff}}}{4}\right)^2 + \omega_m^2\Delta_{\text{eff}}^2 
+ \Delta_{\text{eff}}\frac{u_1 v_2 - u_2 v_1}{m} 
- \frac{\kappa_{\text{eff}}}{2} \frac{u_1 v_1 + u_2 v_2}{m} > 0, \\
s_2 &= \gamma^2 \kappa_{\text{eff}} + \Delta_{\text{eff}}^2 \kappa_{\text{eff}} 
+ \frac{\kappa_{\text{eff}}^3}{4} 
+ \gamma (\kappa_{\text{eff}}^2 + \omega_m^2) 
+ \frac{u_1 v_1 + u_2 v_2}{m} > 0, \\
s_3 &= (\gamma + \kappa_{\text{eff}}) \left( \gamma \kappa_{\text{eff}} + \Delta_{\text{eff}}^2 
+ \frac{\kappa_{\text{eff}}^2}{4} + \omega_m^2 \right)
\left[ \gamma \left(\Delta_{\text{eff}}^2 + \frac{\kappa_{\text{eff}}^2}{4} \right) 
+ \kappa_{\text{eff}} \omega_m^2 - \frac{u_1 v_1 + u_2 v_2}{m} \right] \\
&\quad - \left[ \gamma \left( \Delta_{\text{eff}}^2 + \frac{\kappa_{\text{eff}}^2}{4} \right) 
+ \kappa_{\text{eff}} \omega_m^2 - \frac{u_1 v_1 + u_2 v_2}{m} \right]^2 \\
&\quad - \frac{(\gamma + \kappa_{\text{eff}})^2}{4m} 
\left[ m\omega_m^2( 4\Delta_{\text{eff}}^2 + \kappa_{\text{eff}}^2 )
+ u_1(4\Delta_{\text{eff}} v_2 - 2\kappa_{\text{eff}} v_1) 
- 2u_2(2\Delta_{\text{eff}} v_1 + \kappa_{\text{eff}} v_2) \right] > 0.
\end{aligned}
\label{eq:stability}
\end{equation}
\end{widetext}
The steady-state correlation matrix is obtained as

\begin{equation}
AV + V{A^T} =  - D,
\label{CM}
\end{equation}
where $V$ is the covariance matrix of the system, and the diffusion matrix $D$ are given by

\begin{equation}
D = \left( {\begin{array}{*{20}{c}}
{{\kappa_{\mathrm{eff}}}/2} & 0 & 0 & 0 \\
0 & {{\kappa_{\mathrm{eff}}}/2} & 0 & 0 \\
0 & 0 & 0 & 0 \\
0 & 0 & 0 & {\hbar m\omega_m \gamma \left\{ 2\left[ \exp \left( \frac{\hbar \omega_m}{k_B {\cal T}} \right) - 1 \right]^{-1} + 1 \right\} + \frac{\hbar^2 g_\kappa^2 |c_s|^2}{4\kappa}}
\end{array}} \right).
\label{D}
\end{equation}

\nocite{*}

\bibliography{Supplementary_Materials}